\documentclass[aps,prb,onecolumn,groupedaddress]{revtex4-1}

\pdfoutput=1
\usepackage[utf8]{inputenc}
\usepackage[T1]{fontenc}
\usepackage{hyperref}
\usepackage{graphicx}
\usepackage{siunitx}
\usepackage[version=4]{mhchem}
\usepackage{booktabs}
\usepackage{multirow}

\newcommand{\scro}[0]{\ce{SrCaRu2O6}}
\newcommand{\cro}[0]{\ce{Ca3Ru2O7}}
\newcommand{\mub}[0]{$\mu_\mathrm{B}$}
\newcommand{\vv}[1]{\boldsymbol{\mathrm{#1}}}

\hypersetup{colorlinks = true, citecolor = blue, breaklinks = true}

\begin{document}

\title{Magnetoelectric multipoles in metals}%

\author{Florian Th\"ole}
\email{florian.thoele@mat.ethz.ch}
\affiliation{Materials Theory, ETH Z\"urich, Wolfgang-Pauli-Strasse 27, CH-8093 Z\"urich, Switzerland}

\author{Nicola A. Spaldin}
\email{nicola.spaldin@mat.ethz.ch}
\affiliation{Materials Theory, ETH Z\"urich, Wolfgang-Pauli-Strasse 27, CH-8093 Z\"urich, Switzerland}

\begin{abstract}
We demonstrate computationally the existence of magnetoelectric
multipoles, arising from the second order term in the multipole
expansion of a magnetization density in a magnetic field, in
noncentrosymmetric magnetic metals. While magnetoelectric multipoles
have long been discussed in the context of the magnetoelectric effect in
noncentrosymmetric magnetic \emph{insulators}, they have not previosuly
been identified in metallic systems, in which the mobile carriers screen
any electrical polarization. Using first-principles density functional
calculations we explore three specific systems: First, a conventional
centrosymmetric magnetic metal, Fe, in which we break inversion symmetry
by introducing a surface, which both generates magnetoelectric monopoles
and allows a perpendicular magnetoelectric response. Next, the 
hypothetical cation-ordered perovskite, \scro, in which we study the interplay
between the magnitude of the polar symmetry breaking and that of the
magnetic dipoles and multipoles, finding that both scale proportionally
to the structural distortion. Finally, we identify a hidden \emph{antiferromultipolar} 
order in the noncentrosymmetric, antiferromagnetic metal \cro, and show 
that, while its competing magnetic phases have similar magnetic dipolar 
structures, their magnetoelectric multipolar structures are distinctly 
different, reflecting the strong differences in transport properties.
\end{abstract}
 
\maketitle

\section{\label{sec:level1}Introduction}
The interaction energy, $H_{\mathrm{int}}$ of a magnetization density, $\vv{\mu}(\vv{r})$ with a magnetic field $\vv{H}\left(\vv{r}\right)$ is given in general by the integral of their vector product over all space, 
\begin{equation} \label{eq:inhomog1}
H_{\mathrm{int}} = -  \int \vv{\mu}(\vv{r}) \cdot \vv{H}\left(\vv{r}\right) d^3 \vv{r} \quad .
\end{equation}
The response of many magnetic materials, however, is well described by the approximate interaction
\begin{equation} 
H_{\mathrm{int}} = - \vv{m} \cdot \vv{H}\left(0\right) \quad ,
\end{equation}
where
\begin{equation}
\vv{m} =  \int \vv{\mu} (\vv{r}) d^3 \vv{r}
\end{equation}
is the magnetization and 
$\vv{H}\left(0\right)$ is a \emph{uniform} magnetic field. 
This description captures, for example, the usual Zeeman effect in which the magnetism in a ferro- or ferri-magnetic  tends to align parallel to the field, as well as the well-known susceptibility of antiferromagnets, in which an applied field induces a net magnetization from the compensating magnetic sublattices.

In a particular class of magnetic materials -- those which are insulating and lack a center of inversion symmetry -- this level of treatment is now known to miss important physics, even in the case when the applied field is uniform. 
The simultaneous breaking of space-inversion and time-reversal symmetry in such materials allows them to exhibit the linear magnetoelectric effect, in which an electric field induces a magnetization with magnetoelectric susceptibility $\alpha$ and vice versa\cite{landau1984electrodynamics}. 
This phenomenon is not readily captured by a description of the magnetization at the dipole level, but is revealed transparently in analyses of the next-highest-order multipoles in a multipole expansion of Eqn.~\ref{eq:inhomog1}, since these depend on the product of $\vv{r}$ and $\vv{\mu}(\vv{r})$ and so have the appropriate symmetry \cite{Spaldin2013}. 
Specifically, in the second order of the multipole expansion,
\begin{equation} \label{eq:inhomog2}
H_{\mathrm{int}}^\mathrm{ME} = -  \int r_{i} {\mu}_{j} (\vv{r}) \partial_{i} H_{j}\left(0\right) d^3 \vv{r} \quad ,
\end{equation}
the $ \int r_{i} {\mu}_{j} (\vv{r})d^3 \vv{r}$ component can be decomposed into a sum of three terms,
\begin{align}
a       & = \frac{1}{3} \int \vv{r} \cdot \vv{\mu}(\vv{r}) d^3 \vv{r} \\
\vv{t}  & = \frac{1}{2}  \int \vv{r} \! \times \vv{\mu}(\vv{r}) d^3 \vv{r} \\
q       & = \frac{1}{2} \int \left[r_i \mu_j + r_j \mu_i - \frac{2}{3} \delta_{ij} \vv{r}\! \cdot \vv{\mu(\vv{r})}
\right] d^3\vv{r} \quad .
\end{align}
These are referred to as the magnetoelectric monopole, toroidal moment and quadrupole respectively, and correspond to the diagonal, antisymmetric and symmetric and traceless components of the magnetoelectric tensor \cite{Ederer2007, Spaldin2008, Spaldin2013}. In addition to their connection to the magnetoelectric effect, the magnetoelectric multipoles have proved useful in identifying hidden \textit{antimagnetoelectric} ordering \cite{Spaldin2013}, as well as providing an unambiguous route to defining the size of the local magnetic moment in certain magnetoelectric antiferromagnets \cite{Thole2016}.

Since metallic materials cannot sustain an electric polarization due to screening of the electric field by free charge carriers, the conventional linear magnetoelectric effect can only manifest in insulating materials. Magnetoelectric multipoles, on the other hand, should still be nonzero by symmetry in non-centrosymmetric magnetic metals. 
While higher-order multipoles in metals have recently been considered in the context of the anomalous Hall \cite{Suzuki2016, Suzuki2018} and magnetopiezoelectric effects \cite{Watanabe2017a}, second-order magnetoelectric multipoles have never, to our knowledge, been observed or discussed in the context of metallic systems. 
The purpose of this paper, therefore, is two-fold. First, to compute the properties of some representative magnetic metals with broken inversion symmetry, in order to establish whether magnetoelectric multipoles exist and to determine their magnitudes. And second, to discuss possible properties that might manifest as a result of such a hidden multipolar order.

We proceed by computing the structural and electronic ground states of three model magnetic metals using density functional theory, then extract the atomic-site magnetoelectric multipoles around each ion by transforming the atomic-site density matrix into its irreducible spherical tensor moments \cite{Spaldin2013}. 
We begin by studying a conventional magnetic metal -- iron, Fe -- which is centrosymmetric in its bulk form, and we break the inversion symmetry by creating a surface. 
Provided that the vacuum is sufficiently insulating, the system then becomes insulating perpendicular to the surface, and so can exhibit a linear magnetoelectric effect in the surface normal direction \cite{Rondinelli2008, Duan2008}. To study the influence of the insulator, we also calculate the behavior when the vacuum is replaced by a representative dielectric, MgO. 
In this example we expect magnetoelectric multipoles to occur, and to be substantial only in the vicinity of the interface. 

Next, we investigate the hypothetical non-centrosymmetric magnetic metal, cation-ordered \scro, which has been shown theoretically to have strong coupling between the spins and the polar distortions of the lattice \cite{Puggioni2014, Puggioni2014a, Cao1997a}. 
This system allows us to explore the relationship between the magnitude of the symmetry breaking and the magnitudes of the resulting multipoles by manually modifying the amplitude of the polar distortion.

Finally, we study an established noncentrosymmetric magnetic metal, \cro, in which we anticipate hidden magnetoelectric multipoles in the bulk material because of its combined magnetic order and polar crystallographic structure. Several different magnetic dipole orders are known to exist as a function of temperature \cite{Yoshida2004, Yoshida2005, Bao2008, Baumberger2006, Zhu2016}, allowing us to search for a relationship between dipolar and multipolar magnetic arrangements. 
Additionally, because of its strong spin-charge coupling, \cro\ shows highly anisotropic magnetoresistance\cite{Fobes2011} as well as 2D conductivity at low temperatures; we explore whether these properties can be related to the behavior of the magnetoelectric multipoles.

The remainder of this paper is organized as follows:
In Section \ref{sec:methods}, we describe the computational methods and approximations used.
In Section \ref{sec:results}, we describe in turn our results for Fe surfaces, for \scro\ and for \cro.
We conclude by discussing the relevance of magnetoelectric multipoles for the properties and description of noncentrosymmetric magnetic metals, as well as giving suggestions for further work.


\section{Methods}
\label{sec:methods}
We perform density functional theory (DFT) calculations within the local density approximation (LDA) augmented where appropriate by a Hubbard $U$ correction. Two plane-wave basis codes are employed: For calculations of magnetoelectric multipoles, we use the VASP software package \cite{Kresse1996, Kresse1996a} with projector-augmented wave (PAW) potentials \cite{Kresse1999a}, while for the calculation of field responses, we use the Quantum Espresso package \cite{Giannozzi2009} with ultrasoft pseudopotentials \cite{DalCorso2014}.
 
We model the Fe and Fe/MgO slabs as periodic Fe/vacuum and Fe/MgO/vacuum superlattices, and use an energy cutoff of \SI{540}{eV} and a 8x8x1 Monkhorst-Pack k-point grid\cite{Monkhorst1976}. 
For Mg and O, the $3s$ orbitals and $2s2p$ orbitals, respectively, are treated as valence states in the PAW potentials and ultrasoft pseudopotentials. For Fe, the $3p4s3d$ orbitals are treated as valence states in the PAW potentials, and the $4s3d$ orbitals are treated as valence states in the ultrasoft pseudopotentials.
Electric fields are treated by adding a saw-tooth potential with a compensating dipole layer in the vacuum region\cite{Neugebauer1993, Bengtsson1999a}. 
Magnetic fields are introduced as a Zeeman term in the potential; note that this does not capture contributions from the orbital magnetic response\cite{Bousquet2011a}.

For \scro\ and \cro\ we use energy cutoffs of \SI{550}{eV} and \SI{500}{eV}, and k-point grids of 7x7x5 and 10x10x2 respectively. 
We treat the following orbitals as valence states in the PAW potentials: $3p4s$ for \ce{Ca}, $4s4p5s$ for \ce{Sr}, $4p5s4d$ for \ce{Ru} and $2s2p$ for \ce{O}.
 For \cro, we add a Hubbard $U$ correction \cite{Liechtenstein1995}, $U = \SI{2}{eV}$, on the Ru sites, which is the value used in Ref.~\onlinecite{Zhu2016}. We emphasize that we do not expect DFT to give an accurate description of the detailed electronic structure of these correlated ruthenates, and our emphasis is to reproduce the gross features of \cro\ as a model compound for studying magnetoelectric multipoles.

The atomic-site magnetoelectric multipoles are calculated through the decomposition of the density matrix into irreducible spherical tensor moments, as described in Ref.~\onlinecite{Spaldin2013}.  The dielectric susceptibility was calculated following Giustino et al. \cite{Giustino2005}. All atomically smoothed quantities were calculated by convolution with a trapezoidal kernel whose width was chosen to minimize fluctuations in the bulk-like regions of the slabs \cite{Giustino2005}.
Crystal structure visualizations were produced with VESTA\cite{Momma2011}.


\section{Results and discussion}
\label{sec:results}

\subsection{Fe surfaces and interfaces}

We begin with an analysis of a conventional, centrosymmetric ferromagnetic metal, iron (Fe), in which we break the inversion symmetry by introducing a surface. 
Bulk iron occurs in the bcc structure at low temperatures, with a net magnetization from the ferromagnetic ordering of the local moments on the Fe atoms. The magnetocrystalline anisotropy orients the magnetic moments along a <001> direction, which lowers the space group symmetry to $I4/mm'm'$; the resulting site symmetry of Fe, which is on Wyckoff position $2a$, is $4/mm'm'$. The presence of inversion symmetry at the atomic site forbids the presence of magnetoelectric multipoles. Our first-principles calculations in this setting confirm this fact.

On introduction of a surface, or an interface to a dielectric material, the inversion symmetry is broken and the system becomes insulating in the normal direction. These two properties combine to allow the linear magnetoelectric effect, which has been demonstrated for slabs of other ferromagnetic metals both computationally \cite{Rondinelli2008, Duan2008} and experimentally \cite{Weisheit2007, Zhernenkov2010}.

\begin{figure}[tb]
  \includegraphics{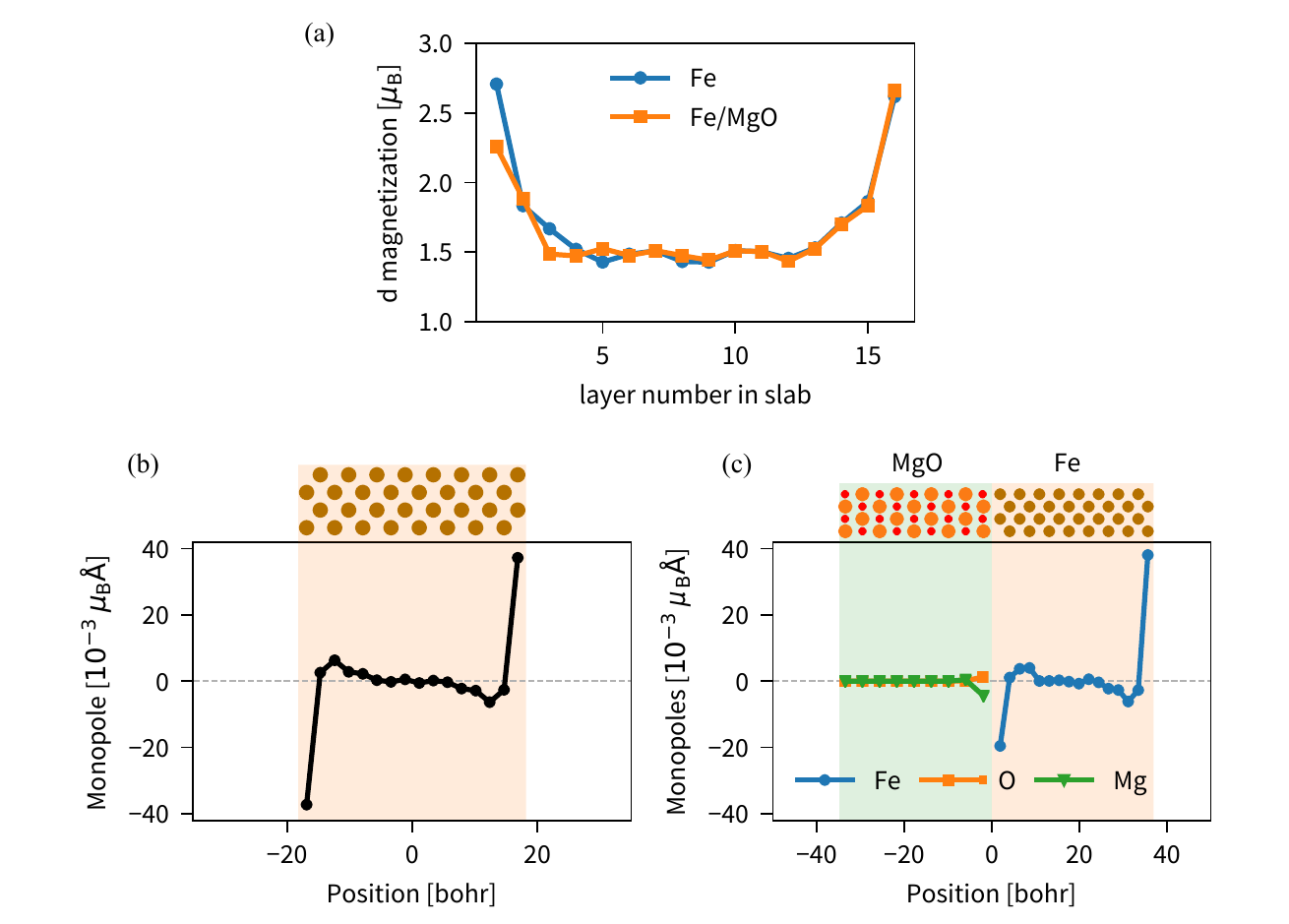}
  \caption{\label{fig:fig1}
  (a) Magnetic dipole moment per Fe atom in the Fe slab (blue) and the Fe/MgO slab (orange). The MgO is at the left side of the figure.
  (b) Magnetoelectric monopoles on each Fe atom in the Fe slab.
  (c) Atomic magnetoelectric monopoles in the Fe/MgO slab.
  }
\end{figure}

Our first model system is a ferromagnetic iron slab, represented by a superlattice containing 16 (001)-oriented layers of bcc-structure Fe separated by \SI{102}{\angstrom}-thick vacuum layers (Fig.~\ref{fig:fig1}(a), upper panel). We set the in-plane lattice constant of the bcc Fe unit cell to $\SI{2.98}{\angstrom}$, corresponding to an in-plane Fe-Fe distance of \SI{2.43}{\angstrom}, so that it forms a coherent interface with MgO in the Fe/MgO heterostructures that we study next  ($a_{\text{MgO}}/\sqrt{2} = \SI{2.98}{\angstrom}$). 
We then relax the out-of-plane Fe-Fe distances, and obtain a Fe-Fe distance of \SI{2.41}{\angstrom} in the outer layers, while the Fe-Fe distance converges to \SI{2.43}{\angstrom} in the bulk-like interior of the slabs. 
For this strain state, the magnetization orients in the uniaxial out-of-plane direction (note that we did not include a demagnetizing field in our calculations); in our presented results it is along the positive direction of Fig.~\ref{fig:fig1}(a) and (b).
In Fig.~\ref{fig:fig1}, we show the magnetic dipole moment on each Fe atom in the slab as a function of the layer.
We obtain a magnetic moment of \SI{2.7}{\mu_B} on the Fe atoms at the surfaces and an interior value of \SI{1.5}{\mu_B}, differing slightly from the LDA bulk value of \SI{2.2}{\mu_B} due to the epitaxial constraint on the in-plane lattice parameters of our slab.

A symmetry analysis reveals that in the slab geometry we are considering --- a tetragonal structure with purely out-of-plane magnetization --- only the magnetoelectric monopole and the $q_z$ component of the magnetoelectric quadrupole are nonzero.
The calculated magnetoelectric monopoles are shown in Fig.~\ref{fig:fig1}(b) as a function of their layer number in the slab. 
(Since the $q_z$ quadrupole component is proportional to the magnetoelectric monopole in our calculation, we do not list it explicitly.)
The monopoles are zero towards the center of the slab, where the local atomic environment is close to the bulk structure and the influence of the inversion-symmetry breaking at the surface becomes negligible, but are nonzero at the surfaces, with opposite signs at opposite surfaces. The opposite signs can be understood either in terms of the opposite position of the vacuum relative to the magnetization orientation at the two surfaces, or by the fact that the surfaces are related to each other by a glide plane, which reverses the monopole sign. 

Next, we repeat our calculations for a heterostructure of the same 16 layers of Fe, this time adjacent to nine layers of (001)-oriented MgO with vacuum on each side. 
In our DFT calculations, the most stable configuration has Fe atoms situated on top of O atoms, as found previously by Butler et al.\cite{Butler2001}, with a Fe-O distance of \SI{2.12}{\angstrom}. (Note that, while our setup is similar to that of Ref.~\onlinecite{Butler2001}, we assume a fixed MgO lattice constant instead of a fixed Fe lattice constant.) As for the Fe slab, we find that the ferromagnetic magnetization in the Fe slab is oriented in the out-of-plane direction. The magnetic moment on the Fe atom at the Fe/MgO interface is reduced to \SI{2.3}{\mu_B} compared with \SI{2.7}{\mu_B} at the Fe/vacuum interface, see Fig.~\ref{fig:fig1}(a).

Our calculated magnetoelectric monopoles are shown in Fig.~\ref{fig:fig1}(c). Again we find that the monopoles are zero in the interior of the Fe slab, and that the surface layers have the largest, and oppositely signed, values. 
The monopole size is reduced to \SI{20e-3}{\mu_B \angstrom} at the Fe/MgO interface, around half of its magnitude at the Fe/vacuum interface, reflecting (although larger than) the decrease of the interfacial Fe magnetic moment from \SI{2.7}{\mu_B} to \SI{2.3}{\mu_B}.
A small monopole is also visible on the Mg and O atoms in the interface layer, indicating a spillover of the magnetic polarization from Fe into the MgO; our calculations yield small magnetic dipole moments of \SI{0.01}{\mu_B} on the Mg atom and \SI{0.06}{\mu_B} on the O atom adjacent to the interface.

The patterns of monopoles that we obtain for both the Fe and Fe/MgO slabs reflect the pattern of electric-field induced magnetization presented for SrRuO$_3$/SrTiO$_3$ heterostructures in Ref.~\onlinecite{Rondinelli2008}, in that they are largest at the surfaces and of opposite sign at either surface of the slab. Therefore as a next step, we calculate the changes in magnetization induced by electric fields applied perpendicular to the surface.

We apply an electric field perpendicular to the slabs by applying a sawtooth potential with a discontinuity in the vacuum region, and set the average field in the supercell to \SI{514}{V/\micro m} for both Fe and Fe/MgO cases. Since the field is screened in the metal, the field in the vacuum depends on the size of the vacuum and dielectric regions in the slab supercell. From our DFT calculations, we find that the electric field in the vacuum region is \SI{626}{V/\micro m} in the Fe slab, and \SI{735}{V/\micro m} in the Fe/MgO slab, where the field is reduced in the dielectric. 
Note that we extract here the electronic contributions to the magnetoelectric response, by performing all calculations at fixed ionic structure.

\begin{figure}[tb]
  \includegraphics{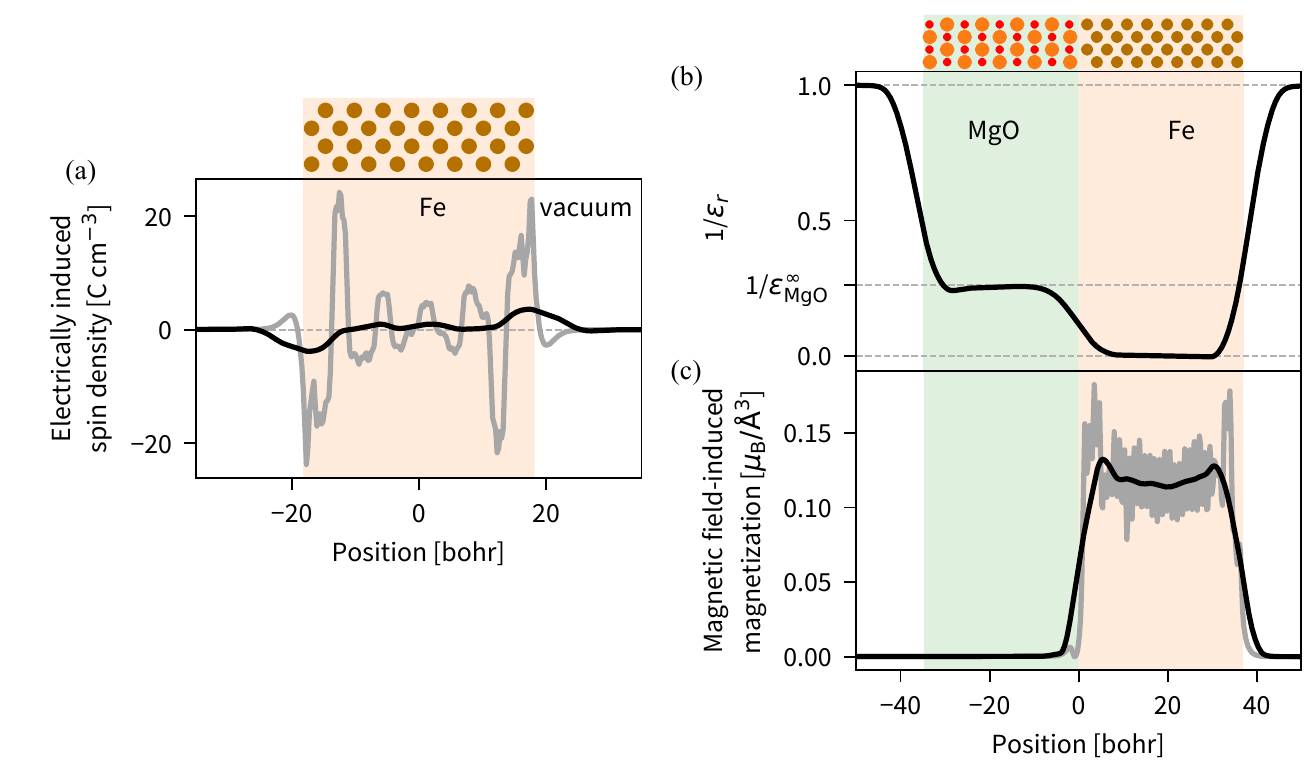}
  \caption{\label{fig:fig2}
  (a) Macroscopically and planar-averaged magnetoelectric response in slab of Fe along the slab axis, i.e., induced magnetization under an electric field. Black line: atomically smoothed response, grey line: unsmoothed response.
  (b) Inverse dielectric constant in the Fe/MgO slab with vacuum on both sides.
  (c) Magnetic-field induced magnetization, that is the magnetic susceptibility, of the Fe/MgO slab.
  }
\end{figure}

The macroscopic and planar-averaged electric field-induced magnetization in the Fe slab is shown in Fig.~\ref{fig:fig2}(a).
We see that the pattern of magnetoelectric response follows closely the pattern of the magnetoelectric monopoles presented above.
First, it is non-zero only in the surface regions, where the inversion symmetry is broken. Second, (and as seen previously for SRO/STO), it is largest on the surface atoms with a small contribution of opposite sign on the next-nearest atoms. Third, at opposite surfaces, the response has opposite signs. Specifically, the change in magnetic dipole moment on the leftmost Fe atom is \SI{-1.18e-3}{\mu_B}, while the change in magnetic dipole moment on the rightmost Fe atom is \SI{1.21e-3}{\mu_B}; in addition, a small overall ferromagnetic component \SI{0.33e-3}{\mu_B} is induced. 
The magnetoelectric response in the Fe/MgO slab follows the same pattern, but with a slight reduction in induced magnetization at the Fe/MgO interface compared to the Fe/vacuum interface, \SI{1.33e-3}{\mu_B}. Note however that due to our choice of electrostatic boundary conditions, the voltage at each interface is different, which forbids a direct comparison of the magnitude of the induced response.
Again there is a net induced ferromagnetic component of \SI{0.38e-3}{\mu_B}.

Since the magnetoelectric susceptibility, $\alpha$, is a bulk property it is not the relevant quantity for describing surface electric-field induced magnetism. Instead, Rondinelli et al. introduced the concept of spin capacitance, $C^s = \frac{\sigma^s}{V}$, which, by analogy to the usual charge capacitance, $C$, is the spin polarization per unit area induced by the voltage $V$ \cite{Rondinelli2008}.
They then suggested a magnetoelectric ``figure of merit'' given by $\eta=C^s / C $ \cite{Rondinelli2008}. For our Fe/vacuum slabs, we obtain $\eta=0.28$, similar to that found earlier for the SRO/STO interface \cite{Rondinelli2008}, and for the Fe/MgO interface in our Fe/MgO slabs, we obtain $\eta=1.27$. 
The large and somewhat unintuitive $\eta > 1$ is a result of spin transfer from the majority to minority channel at the Fe/MgO interface, in addition to the capacitive charge and spin accumulation. Whether this huge magnetoelectric figure of merit is revelant for the favorable tunneling magnetoresistance in Fe/MgO heterostructures \cite{Butler2001} is an interesting question for future exploration.

Lastly, we reflect on the relation between susceptibility and magnetoelectric response. 
In bulk materials, the diagonal magnetoelectric response $\alpha$ is known to be related to the size of the magnetoelectric monopole per unit volume by 
\begin{equation}
  \alpha = c (\epsilon_r-1) \chi_m A \;,
  \label{eq:free-energy}
\end{equation}
where $\epsilon_r$ is the relative permittivity, $\chi_m$ is the magnetic susceptibility, $A$ is the  magnetoelectric monopole per unit volume and $c$ is a proportionality constant \cite{Spaldin2013,Thole2016}. 

In Fig.~\ref{fig:fig2}(b) we show our calculated inverse relative permittivity and in Fig.~\ref{fig:fig2}(c) the magnetization induced by a Zeeman field of \SI{1}{mT} for the Fe/MgO slab. (The interface in the Fe/vacuum slab behaves similarly to the Fe/vacuum interface shown here). 
As expected, the relative permittivity is unity in the vacuum and diverges in the metallic region. It has a finite value in the dielectric and at the interfaces of the metal. 
The magnetic susceptibility, on the other hand, is only non-zero in the metallic (Fe) region. 
We see that the product of both susceptibilities is non-zero and finite only in the region in which we observe a magnetoelectric response.

In summary, our first-principles calculations confirm that magnetoelectric monopoles and quadrupoles can be generated in nominally centrosymmetric Fe by introducing interfaces that break the inversion symmetry. These magnetoelectric multipoles are large only at the interfaces and vanish rapidly towards the bulk region. At the same time, since such a system is insulating in the direction normal to the surface, it exhibits a magnetoelectric response, which, like the multipoles, is large only at the interfaces. We find that the sizes of the magnetic dipole moment and the magnetoelectric monopole depend on the detailed nature of the interface, with those at the Fe/MgO interface being smaller than at the Fe/vacuum interface. Finally, we show that, while the magnetoelectric response coincides with the region in which the product of electric and magnetic susceptibilities is finite and non-zero, there is no obvious connection between their magnitudes.

\subsection{\ce{SrCaRuO6}}
\label{sec:scro}
We now turn our attention to the hypothetical magnetic polar metal, A-site ordered \scro, in which it has been shown computationally that the magnetism can be modified by modulation of the non-centrosymmetric structural distortion. 
The disordered solid solution series \ce{Sr_{1-x}Ca_{x}RuO3} of the isostructural perovskites \ce{SrRuO3} and \ce{CaRuO3} exists experimentally, and is metallic and non-polar at all compositions, with the centrosymetric $Pnma$ space group  
(octahedral tilt pattern $a^{+}b^{-}c^{-})$\cite{Eom1992}.
From $x=0$ up to $x \approx 0.7$, it is an itinerant ferromagnet with $T_\mathrm{C} \approx \SI{57}{K}$\cite{Cao1997a}. Above this value of $x$, the ferromagnetic order is suppressed.

\begin{figure}[tb]
  \includegraphics[width=1\textwidth]{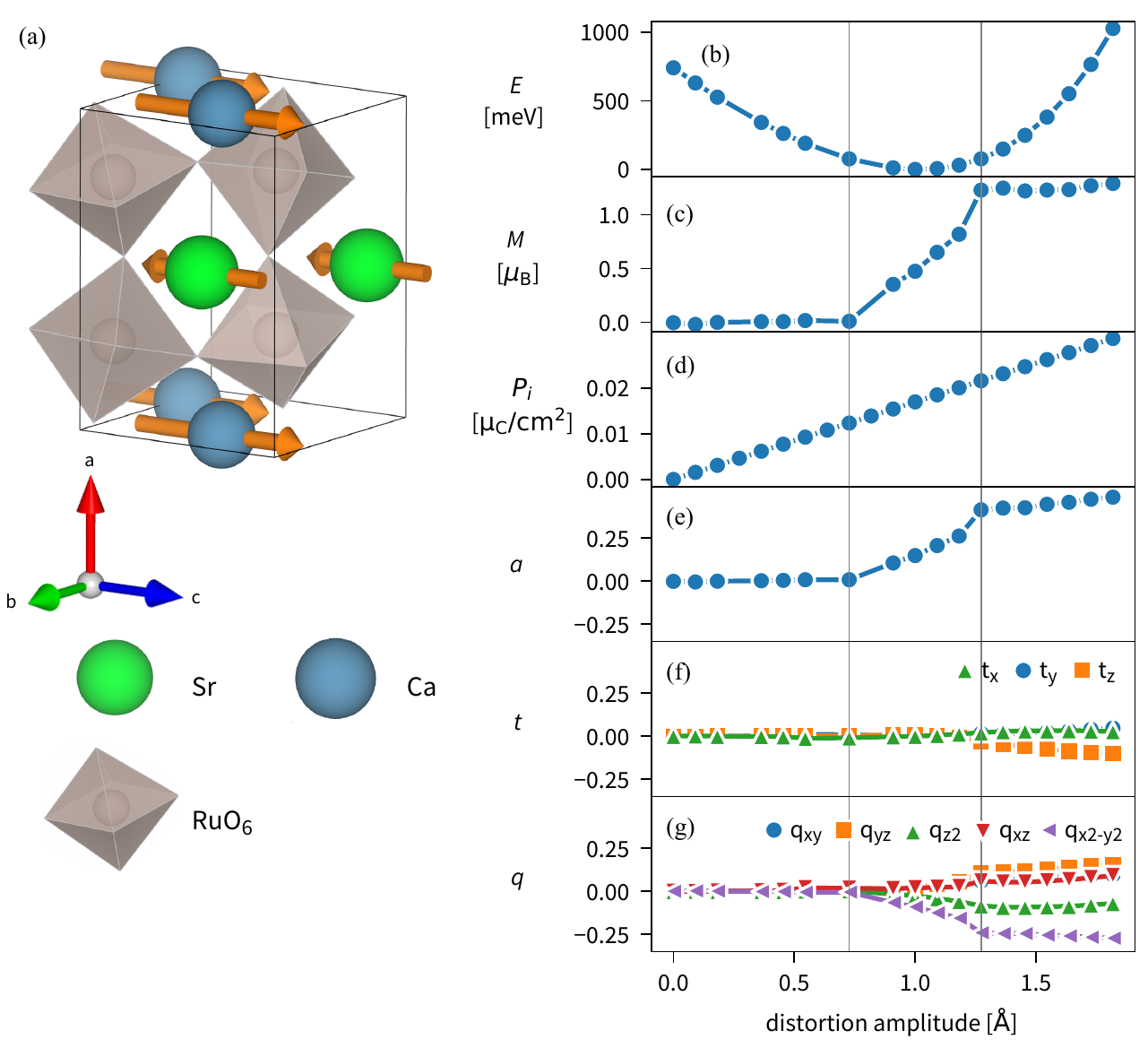}
  \caption{\label{fig:scro-gamma5}
  (a) Structure of A-site cation-ordered \ce{SrCaRu2O6}. Orange arrows denote the atomic displacements in the polar $\Gamma_5^-$ mode of \scro. 
  (b-g) Properties of \ce{SrCaRu2O6} as a function of polar mode amplitude:
  (b) Energy, $E$ (c) ferromagnetic magnetization, $M$ (d) ionic polarization of the lattice, $P_i$ (e) magnetoelectric monopoles, $a$ (f) all components of the toroidal moment, $t$ (g) all quadrupoles, $q$.
   Note that the normalization of the distortion mode in this work is such that the relaxed structure has an amplitude of \SI{1}{\angstrom}, which is different from the normalization used in Ref.~\onlinecite{Puggioni2014}.}
\end{figure}

Puggioni et al.\ showed using DFT calculations that inversion symmetry is broken when the A-site cations are ordered in layers along the $c$ direction \cite{Puggioni2014}, while the magnetic and metallic behavior persist. Constraining the $a$ and $b$ lattice constants to be equal to mimic coherent epitaxial growth, they obtained a $Pmc2_1$ space group, with the same tilt pattern as the $Pnma$ of the disordered alloy and an additional distortion corresponding to the polar $\Gamma_5^-$ mode (shown in Fig.~\ref{fig:scro-gamma5}(a)) of the parent $P4/mmm$ space group. Interestingly, suppression of the polar $\Gamma_5^-$ mode caused a collapse of the ferromagnetic order \cite{Puggioni2014}; we will use this fact to investigate the interplay between polarization, magnetization and magnetoelectric multipoles.

Using the ground-state structure of the layered A-site ordered compound from Ref.~\onlinecite{Puggioni2014}, we calculate the electronic and magnetic structure with spin-orbit coupling included.
Within the local density approximation, we obtain a magnetic moment on the Ru atoms of \SI{0.1}{\mu_B}; this increases to \SI{0.7}{\mu_B} with even a small $U=\SI{0.5}{eV}$. Since the experimentally measured value for the magnetic moment per Ru atom in disordered \ce{Sr_xCa_{1-x}RuO3} with $x=0.53$ is below 0.2 \mub. \cite{Cao1997a}, we do not apply a Hubbard $U$ correction in the following.  
 We find that the magnetic dipole moments are oriented in the orthorhombic $b$ direction, leading to the magnetic space group $Pm'c2_1'$. Since the site symmetry of the Ru atoms on Wyckoff position $4c$ in this space group is $1$, all magnetoelectric multipoles are allowed on each site.
 Symmetry analysis of the allowed arrangements of the magnetoelectric multipole orders in this spacegroup yields the results summarized in Tab.~\ref{tab:ru-sym-order}, where + and - indicate the sign of the allowed multipoles. 
 For the $t_x$ toroidal moment and the $q_{yz}$ quadrupole, a ferro-type order is allowed, while the remaining multipoles order in different antiferro-type patterns. 
 If the system were insulating, the corresponding bulk magnetoelectric effect in an insulator would have two nonzero components $\alpha_{23}$ and $\alpha_{32}$, with the symmetric part, $(\alpha_{23} + \alpha_{32})/2$, determined by the magnetoelectric quadrupole $q_{yz}$, and the antisymmetric part, $(\alpha_{23} - \alpha_{32})/2$, determined by the toroidal moment $t_x$.

\begin{table}[tb]
  \begin{ruledtabular}
\begin{tabular}{cccccc}
Atom  & $m_y$ & $a$, $q_{z^2/x^2-y^2}$ & $t_x$,$q_{yz}$  & $t_y$, $q_{xz}$ & $t_z$, $q_{xy}$ \\
\hline
Ru1  & + & + & + & + & - \\
Ru2  & + & - & + & - & - \\
Ru3  & + & - & + & + & + \\
Ru4  & + & + & + & - & + \\
\end{tabular}
\end{ruledtabular}
\caption{\label{tab:ru-sym-order}Symmetry analysis of the dipolar and multipolar order on the Ru atoms, which occupy the Wyckoff position $4c$ in the $Pmc2_1$ space group. The Ru magnetic dipole moments order ferromagnetically and are oriented along $y$, leading to the magnetic space group $Pm'c2_1'$. This results in a ferro ordering of the $t_x$ and $q_{xy}$ multipoles, and the antiferro orderings shown for the other multipoles.}
\end{table}

\begin{table}[tb]
  \begin{ruledtabular}
\begin{tabular}{rrrrrrrrr}
\multicolumn{9}{c}{[\SI{e-3}{\mu_B \angstrom}]} \\
$a$ & $t_x$ & $t_y$ & $t_z$ & $q_{xy}$ & $q_{yz}$ & $q_{z^2}$ & $q_{xz}$ & $q_{x^2-y^2}$\\
\hline
 0.15 &     -0.00 &      0.00 &      0.01 &      0.01 &      0.03 &     -0.03 &      0.02 &     -0.09 \\
\end{tabular}
\end{ruledtabular}
\caption{\label{tab:ru-multipoles} Calculated size of the magnetoelectric multipoles in \scro\ in the equilibrium structure.}
\end{table}

Next, we analyze the behaviour of the magnetic order when the polar mode amplitude is changed.
Keeping all modes except for the polar mode at their bulk amplitudes, we calculate the energy and magnetic structure for several different amplitudes of the polar mode between zero and \SI{1.8}{\angstrom}, where we normalize the mode amplitude such that the ground state structure corresponds to an amplitude of \SI{1}{\angstrom}.
Figs.~\ref{fig:scro-gamma5}(b) and (c) show our calculated energy and magnetization as a function of amplitude. 
At zero and small amplitude, we obtain a nonmagnetic solution, with the onset to the ferromagnetic state occurring at a mode amplitude of \SI{0.72}{\angstrom}. The ferromagnetic moment then increases up to \SI{1.2}{\mu_B} at a mode amplitude of \SI{1.27}{\angstrom} where it saturates. The ground-state structure has its mode amplitude in the intermediate region in which the magnetization has not reached its saturation value. Note that at all amplitudes, antiferromagnetic orders with the same magnetic unit cell are higher in energy than the calculated non-magnetic and ferromagnetic orders.

Next, we compute the magnetoelectric multipoles as a function of the polar mode amplitude and show the size of the monopole, toroidal moment and quadrupole components as a function of mode amplitude in Fig.~\ref{fig:scro-gamma5}(c). 
As expected, at zero and small distortion amplitude there are no multipoles, since there is no time-reversal symmetry breaking magnetic dipole order. 
In the intermediate region, the size of all multipole components increases roughly proportionally to the size of the ferromagnetic dipole moment. The sizes of the multipoles at the ground state amplitude are listed in Tab.~\ref{tab:ru-multipoles}. At this amplitude, the toroidal moments are essentially zero, and the magnetoelectric monopole is the largest component.
In the region in which the magnetization saturates, the magnetoelectric multipoles saturate also, showing negligible change with further increase of the polar mode amplitude. 

In summary, we find a region of polar distortion amplitude in which both the usual dipolar magnetization as well as the magnetoelectric multipoles scale proportionally to the strength of the inversion-symmetry breaking. While the polar mode amplitude can not be modified using an electric field due to metallic screening, the reciprocal effect -- modification of the magnetization with a magnetic field to tune the amplitude of the polar structural distortion -- should be accessible. This hidden magnetoelectric response might be observable using second harmonic generation, which is sensitive to the inversion symmetry breaking. Since the magnetoelectric multipoles saturate together with the magnetization even when the polarization amplitude continues to increase, we conclude that they scale with the magnitude of the dipolar magnetization rather than being explicitly sensitive to the magnitude of the inversion symmetry breaking.


\subsection{\ce{Ca3Ru2O7}}
\begin{figure}[bt]
  \includegraphics[width=1.0 \textwidth]{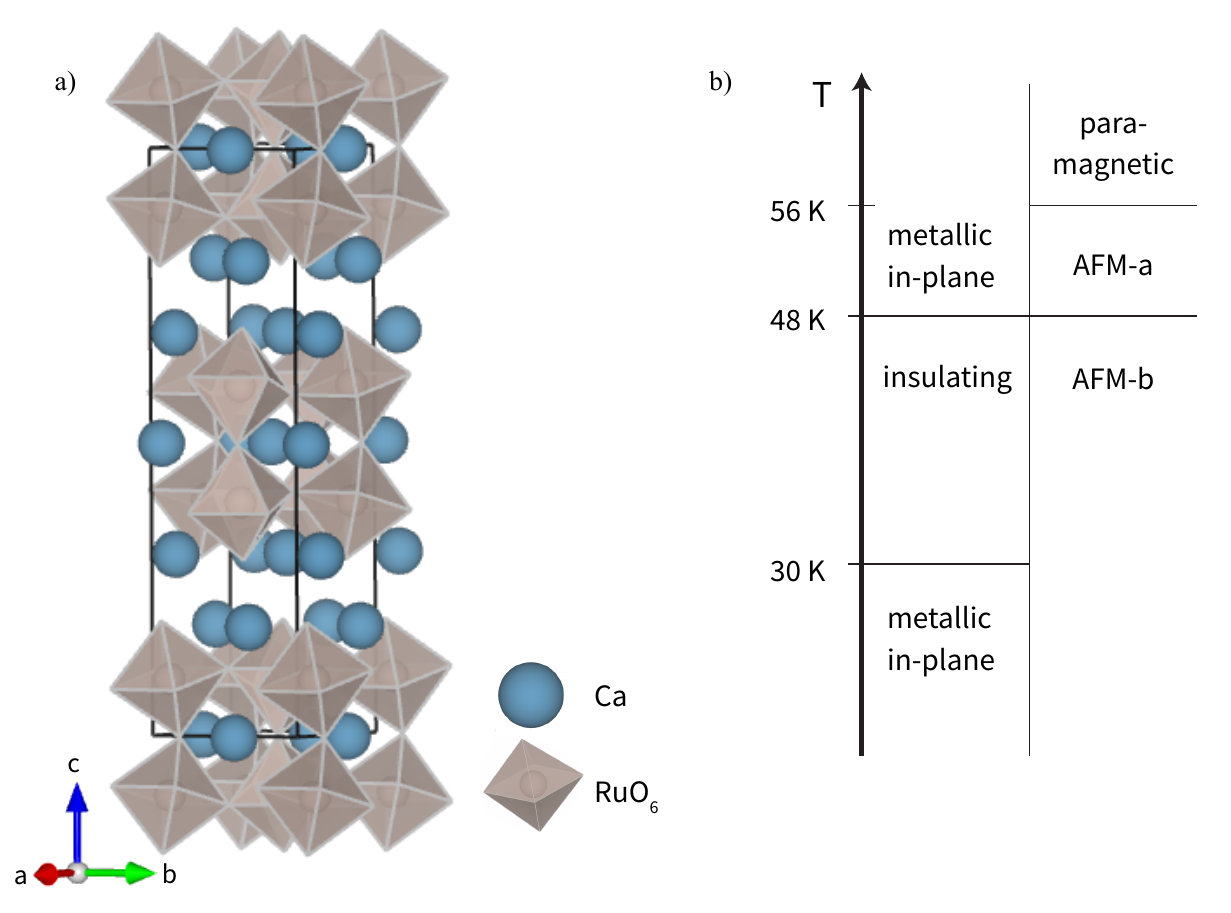}
  \caption{\label{fig:cro-structure}
  (a) Crystal structure of \cro.
  (b) Schematic of the electronic and magnetic phase diagram.
  }
\end{figure}

\cro\ is a member of the \ce{A_{n+1}B_nX_{3n+1}} Ruddelsden-Popper series with $n=2$; its structure is shown in Fig.~\ref{fig:cro-structure}(a). 
Much previous work has focused on understanding its electrical and magnetic phase diagram, sketched in Fig.~\ref{fig:cro-structure}(b)\cite{Yoshida2005, Yoshida2004, Baumberger2006}.
Above $T_\mathrm{MIT}=\SI{48}{K}$, the system is metallic in-plane and insulating out-of-plane, in the sense that the in-plane resistivity $\rho_a$ decreases with decreasing temperature, while the out-of-plane resistivity $\rho_c$ shows increasing resistivity with decreasing temperature\cite{Yoshida2004}. Below $T_\mathrm{MIT}$, it shows insulating behaviour in all directions, until, below \SI{30}{K}, it regains metallic conductivity in the a-b plane. 
 At $T_\mathrm{N}=\SI{56}{K}$, within the metallic phase, it undergoes a phase transition from paramagnetic to antiferromagnetic with the so-called AFM-$a$ arrangement, in which each double layer is ordered ferromagnetically with AFM coupling between the double layers. 
In this phase, the anisotropy causes the spins to lie along the $a$ axis\cite{Bao2008}.
 At $T_\mathrm{MIT}$, the magnetic moments reorient to align parallel to the $b$ axis while keeping the overall magnetic ordering, forming the so-called AFM-$b$ state. 
The c-axis magnetoresistance is different between the two orientations of the AFM order and exhibits a pronounced temperature dependence around the transition temperatures\cite{Fobes2011}. This has been attributed to the strong spin-charge coupling in the material\cite{Peng2016, Baumberger2006}.

In the following, we will show that, while the point group symmetry of the material contains time reversal,  magnetoelectric multipoles are realized in both antiferromagnetic phases because the \emph{site symmetry} is not time-reversal symmetric.
We will further show that magnetoelectric multipoles provide a sensitive indicator of the differences in hybridization between the two different magnetic orientations, and so can be helpful in revealing the coupling between charge and spin degrees of freedom. 
As a consequence, we will argue that the magnetoelectric multipoles are a sensitive tool for characterizing the nature of the microscopic magnetic anisotropies, even when the magnetization densities are essentially identical for different choices of easy axis.

We start by analyzing the magnetic symmetry of the system.
The $b$-axis anisotropy of the $Bb2_1m$ space group\cite{Yoshida2005, Bao2008} leads to a magnetic space group of $B_Pb'2_1m'$ (in OG setting), while the AFM-$a$ phase has the magnetic space group $B_Pb2_1'm'$. 
In both cases, the magnetic point group is $mm21'$ and so contains time-reversal symmetry. 
This prohibits a ferro-type order of magnetoelectric multipoles, so if Ca$_3$Ru$_2$O$_7$ were insulating a bulk magnetoelectric effect would not be allowed. Antiferro-type multipolar orders are allowed, however, because the time-reversal symmetry in this magnetic space group occurs in combination with a translation through the $B$ centering of the unit cell.
This is consistent with the $1$ site symmetry of the Ru atoms on Wyckhoff position $8b$, which does not contain time-reversal (or any other) symmetry. As a result, magnetoelectric multipoles occur on the individual Ru sites, arranged in the antiferro-type orders shown in Tab.~\ref{tab:cro-symmetry}.

\begin{table}[tb]
  \begin{ruledtabular}
  \begin{tabular}{c|cccc|cccc}
\multirow{2}{*}{Wyckoff position $8b$} & \multicolumn{4}{c|}{AFM-a} & \multicolumn{4}{c}{AFM-b}  \\
    & $a$, $q_{z^2/x^2-y^2}$ & $t_x$, $q_{yz}$  & $t_y$, $q_{xz}$ & $t_z$, $q_{xy}$ 
      & $a$, $q_{z^2/x^2-y^2}$ & $t_x$, $q_{yz}$  & $t_y$, $q_{xz}$ & $t_z$, $q_{xy}$ \\
\hline
($x$, $y$, $z$) 	 & + 	  & - 	  & + 	  & - 	  & - 	  & - 	  & - 	  & - 	 \\
($-x$, $y + 1/2$, $-z$) 	 & - 	  & - 	  & - 	  & - 	  & - 	  & + 	  & - 	  & + 	 \\
($-x$, $y + 1/2$, $z$) 	 & - 	  & + 	  & + 	  & - 	  & - 	  & - 	  & + 	  & + 	 \\
($x$, $y$, $-z$) 	 & + 	  & + 	  & - 	  & - 	  & - 	  & + 	  & + 	  & - 	 \\
($x + 1/2$, $y$, $z + 1/2$) 	 & - 	  & + 	  & - 	  & + 	  & + 	  & + 	  & + 	  & + 	 \\
($-x + 1/2$, $y + 1/2$, $-z + 1/2$) 	 & + 	  & + 	  & + 	  & + 	  & + 	  & - 	  & + 	  & - 	 \\
($-x + 1/2$, $y + 1/2$, $z + 1/2$) 	 & + 	  & - 	  & - 	  & + 	  & + 	  & + 	  & - 	  & - 	 \\
($x + 1/2$, $y$, $-z + 1/2$) 	 & - 	  & - 	  & + 	  & + 	  & + 	  & - 	  & - 	  & + 	 \\
\end{tabular}
\end{ruledtabular}
\caption{\label{tab:cro-symmetry}
Allowed multipole orders on Wyckhoff position $8b$ in the $Bb2_1m$ space group, for antiferromagnetic arrangements of magnetic moments aligned along the $a$ axis (AFM-$a$) and along the $b$ axis (AFM-$b$).
}
\end{table}

The structural and magnetic properties of Ca$_3$Ru$_2$O$_7$ have been investigated previously using density functional calculations \cite{Singh2006, Ke2011, Zou2016, Zhu2016, Liu2011b}, and the calculated properties were found to depend strongly on the choice of Hubbard $U$.
In all calculations, the magnetic ground state was found to have the magnetic ordering of the AFM-$b$ phase. Without including a Hubbard $U$ or spin-orbit interactions, calculations using LDA or GGA obtained a metallic system\cite{Singh2006, Ke2011}.
Including a moderate $U=\SI{2}{eV}$ on the Ru atoms led to a gap opening in higher energy AFM phases, but the AFM-$b$ phase retained its metallic character \cite{Zou2016}. Further increase of $U$ to \SI{3.5}{eV} and inclusion of spin-orbit interactions opened a gap in the AFM-$b$ phase \cite{Liu2011b}. We emphasize that the most appropriate description of this correlated oxide is a difficult and ongoing question \cite{Deng2016} which we do not address here. Rather, since our motivation is to use Ca$_3$Ru$_2$O$_7$ as a model system to establish the existence of magnetoelectric multipoles, we take the simplest method that gives qualitatively correct behavior, that is the LDA method with $U=\SI{2}{eV}$.

Using the experimental crystal structure from Ref.~\onlinecite{Yoshida2005}, we calculate the electronic structure as described in Sec.~\ref{sec:methods}, imposing the AFM-$b$ and AFM-$a$ magnetic order in turn. In both cases we obtain a metallic system in which the density of states at the Fermi level stems mainly from  the Ru $d$ orbitals, with the AFM-$a$ phase being \SI{1.5}{meV/Ru} atom higher in energy.
We find that a small antiferromagnetic tilting of the magnetic moments of the Ru atoms away from the $a$ (for AFM-$a$) or $b$ (for AFM-$b$) easy axis is energetically favorable; this is allowed by symmetry for the Wyckoff position $8b$ occupied by the Ru atoms. In the following, we neglect this small rotatation and constrain the moments to lie along the $a$ axis for the AFM-$a$ and along the $b$ axis for the AFM-$b$ structure, respectively, to allow for a more straightforward comparison. 

\begin{figure}[bt]
  \includegraphics[width=1.0 \textwidth]{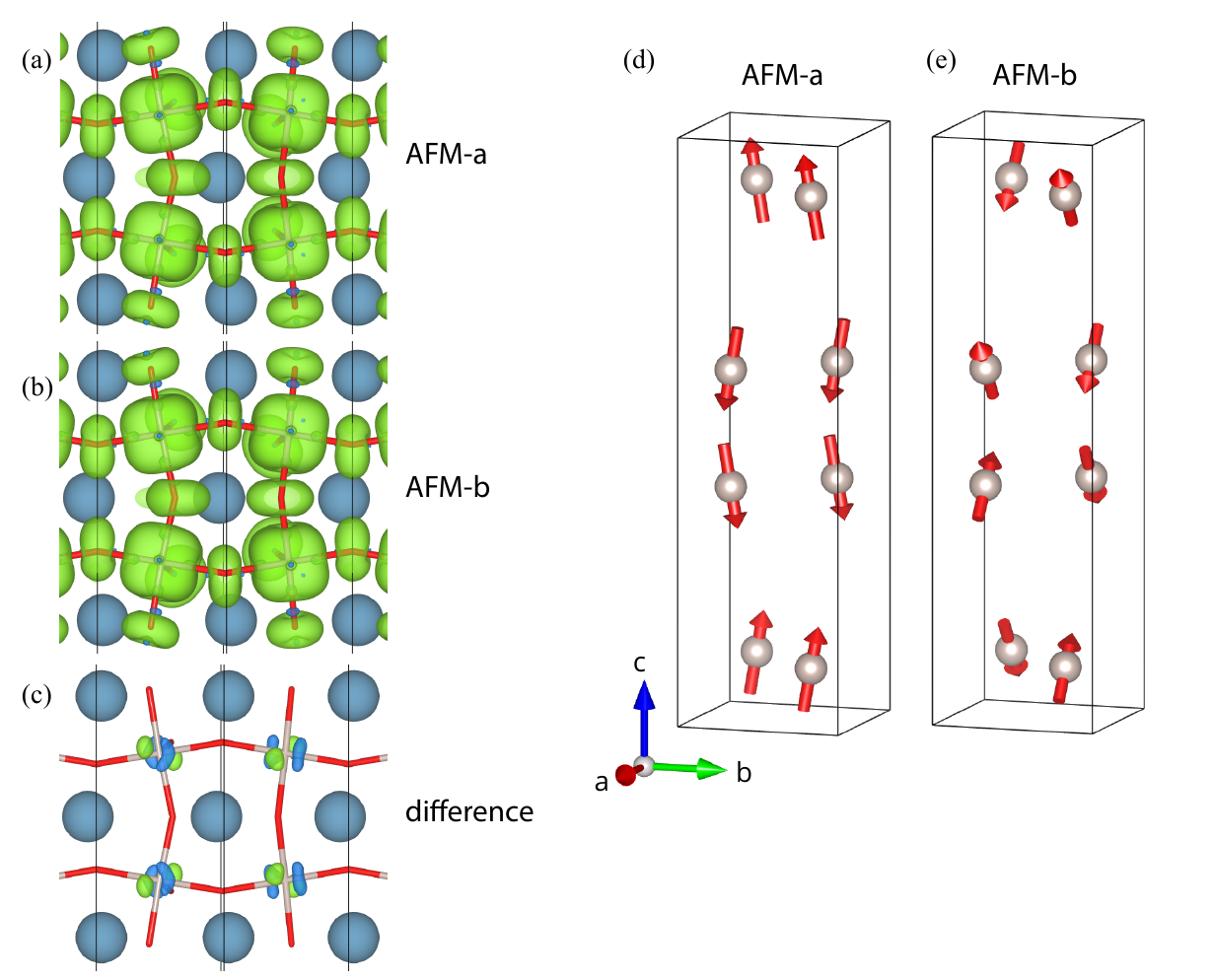}
  \caption{\label{fig:cro-tmoms}
  (a): Magnetization density for AFM-$a$-ordered \cro. The isosurface level is \SI{2e-3}{\mu_B/\angstrom^3}.
  (b): Magnetization density for AFM-$b$-ordered \cro. The isosurface level is \SI{2e-3}{\mu_B/\angstrom^3}.
  (c): Difference between magnetization densities of AFM-$a$ and AFM-$b$ orderings. The isosurface level is \SI{1e-3}{\mu_B/\angstrom^3}.
  (d), (e): Toroidal moments on Ru atoms in \cro\ when the magnetic moments are aligned along $a$ (AFM-$a$) or $b$ (AFM-$b$). 
  For the AFM-$a$ structure, the toroidal moments lie in the $b-c$ plane and are approximately oriented along $c$, while for AFM-$b$, the toroidal moments lie in the $a-c$ plane and are approximately oriented along $c$.
  The \ce{Ca} and \ce{O} atoms are not shown.
  }
\end{figure}

\begin{figure}[bt]
  \includegraphics{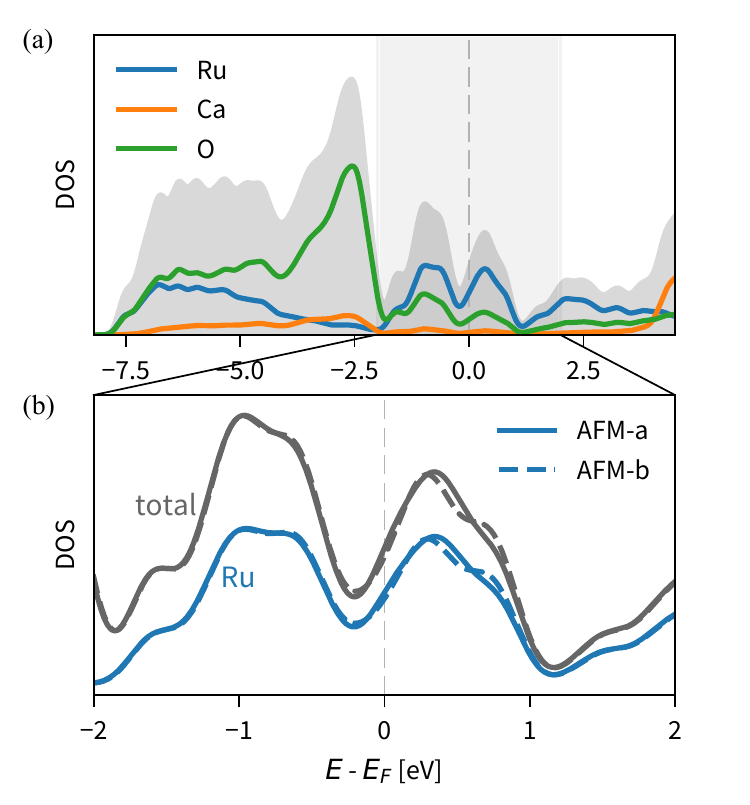}
  \caption{\label{fig:cro-dos}
  (a) Atomic-orbital projected density of states (DOS) for \cro\ with AFM-$a$ order. (b) Comparison of the DOS around the Fermi energy for AFM-$a$ and AFM-$b$ orders.
  }
\end{figure}

The magnetization density along the $a$ direction for the AFM-$a$ phase is shown in
Fig.~\ref{fig:cro-tmoms}(c), while the magnetization density along the $b$ direction for the AFM-$b$ phase is shown in Fig.~\ref{fig:cro-tmoms}(d). In spite of the striking difference in properties measured for the two phases
--- AFM-$b$ has a higher out-of-plane resistivity with a different magnetic-field dependence at temperatures around the metal-insulator transition ---, we see that the shape of the magnetization densities is indistinguishable on this scale.
Even the difference density, shown in Fig.~\ref{fig:cro-tmoms}(d), is tiny, although one can resolve small changes in the regions close to the Ru atoms, This suggests a small rehybridization of the Ru~$d$ orbitals, even though this is barely visible in the calculated density of states of the Ru $d$ bands around the Fermi level, shown in Fig.~\ref{fig:cro-dos}.

These small differences in magnetization are revealed much more strikingly in the magnetoelectric multipoles, which we report in Tab.~\ref{tab:cro-comparison-tmoms}. As a result of the low symmetry, all components are nonzero in both magnetic phases. 
We see that the monopole term is strongest in the AFM-$b$ phase, while the average toroidal and quadrupole moments are larger in the AFM-$a$ phase.

\begin{table}[bt]
\begin{ruledtabular}
\begin{tabular}{lrrrrrrrrr}
 & \multicolumn{9}{c}{($10^{-3} \mu_{\mathrm{B}}\AA$)} \\
 &     a  
 &     $t_x$ & $t_y$ & $t_z$  
 &     $q_{xy}$ & $q_{yz}$ & $q_{z^2}$ & $q_{xz}$ & $q_{x^2 - y^2}$ \\
\hline
AFM-a & 0.29 & -0.01 & 0.06 & -0.35 & -0.36 & 0.01 & -0.42 & -0.19 & 0.43 \\
AFM-b & -0.60 & -0.19 & 0.00 & -0.25 & 0.41 & -0.06 & 0.16 & 0.04 & 0.36 \\
\end{tabular}
\end{ruledtabular}
\caption{\label{tab:cro-comparison-tmoms}Sizes of the Ru-atom magnetoelectric multipoles in \cro, for both AFM$-a$ and AFM$-b$ phases.}
\end{table}

We focus on the toroidal moments, shown in Figs.~\ref{fig:cro-tmoms}(d) and (e), to analyze the differences in multipole behavior between the two magnetic orderings. 
Since there is no component of toroidal moment parallel to the magnetic moment, the toroidal moments are aligned perpendicular to the magnetic moments, that is in the $b$--$c$ plane for AFM-$a$ and the $a$--$c$ plane for AFM-$b$. They are tilted away from the $c$ axis by \ang{10} in AFM-$a$ and \ang{36} in AFM-$b$. It is clear that the crystallographic differences in the orthorhombic $a$ and $b$ directions, which causes the tiny changes in magnetizations from the  different hybridization of the Ru~$d$--O~$p$ orbitals manifest as distinctly different toroidal moments with different magnitudes and relative orientations.

In summary, while the symmetry of \cro\ does not allow macroscopic multipole order, we find a hidden antiferromultipolar order in the magnetoelectric monopole, toroidal moment and quadrupole.
The distinctly different magnetoelectric multipoles displayed by different magnetic orientations provide a useful handle for quantifying subtle rearrangements of magnetization density that are not readily revealed from an analysis of the magnetic dipole contribution alone. 
While there is no obvious connection between the differences in magnetoelectric multipoles and the differences in transport properties between the two magnetic orderings, this could be an interesting consideration for future work. 
 
  
\section{Conclusion}
In conclusion, we have shown for the first time that second-order magnetoelectric multipoles exist as a ``hidden order'' in non-centrosymmetric magnetic metallic systems. We investigated their behaviour in three different systems: the surface of ferromagnetic Fe, which is centrosymmetric in its bulk form, the hypothetical polar magnetic metal \scro, in which the polar mode strongly modifies the magnetization, and the known noncentrosymmetric antiferromagnetic metal, \cro. 

We identified magnetoelectric monopoles at surfaces and interfaces of centrosymmetric Fe, and showed that they are consistent with the carrier-mediated electric-field induced magnetism previously reported in related systems. We showed that the magnetoelectric multipoles can be controlled by modulating the amplitude of the polar $\Gamma_5^-$ mode in \scro, with their size corresponding closely to that of the corresponding dipolar magnetization. Since the amplitude of the polar mode can not be modified by an applied electric field, however, their is no accompanying magnetoelectric effect. Finally, we identified hidden anti-ferro-ordered magnetoelectric multipoles in both magnetic phases of \cro, and showed that they depend strongly on the orientation of the antiferromagnetic dipole moments. Consequently, they provide a sensitive indicator of the changes in magnetization density associated with the spin reorientation.

We hope that the identification of magnetoelectric multipoles in magnetic metals achieved in this work motivates future studies of the relationship between the magnetoelectric multipoles and properties such as magnetoeresistance and spin-dependent transport in polar magnetic metals.

\section{Acknowledgements}
This work was supported financially by the ETH Z\"urich, by the Max R\"ossler Price of the ETH Z\"urich, the K\"orber Foundation and by the Sinergia program of the Swiss National Science Foundation Grant No. \mbox{CRSII2\_147606/1}. This work was supported by a grant from the Swiss National Supercomputing Center (CSCS) under project IDs s624 and p504.
We thank Massimiliano Stengel, Maxim Mostovoy and Sang-Wook Cheong for insightful discussions.

\bibliography{thelibrary}

\end{document}